\documentclass[11pt]{article}
\usepackage[english]{babel}
\renewcommand{\thefootnote}{\alph{footnote}}
\usepackage{etoolbox}
\usepackage{verbatim}

\makeatletter
\renewcommand*{\@fnsymbol}[1]{\ifcase#1\or1\else\@arabic{\numexpr#1-1\relax}\fi}

\makeatother
\usepackage{microtype} 
\usepackage[numbers,sort&compress]{natbib}
\usepackage{graphicx} 
\usepackage{amsmath}
\usepackage{amsthm} 
\usepackage{amsfonts} 
\usepackage{amssymb} 
\usepackage{fullpage} 
\usepackage{color} 
\usepackage{latexsym} 
\usepackage{graphicx}
\usepackage{wrapfig}
\usepackage{blindtext}
\usepackage{subfigure} 
\usepackage{soul} 
\usepackage{enumerate} 
\usepackage[nameinlink]{cleveref} 
\usepackage{thm-restate} 
\usepackage{xfrac} 
\usepackage{mathtools,mathdots}
\usepackage{physics}
\usepackage{verbatim}
\usepackage{algorithmic}
\usepackage{algorithm}
\usepackage{appendix}

\usepackage{diagbox} 
\usepackage{makecell} 
\usepackage{array,booktabs} 
\usepackage{hhline} 
\usepackage[table]{xcolor} 
\usepackage{multirow} 

\newtheorem{theorem}{Theorem}[section] 
 
\newtheorem{lemma}[theorem]{Lemma}

\newtheorem{proposition}[theorem]{Proposition}

\theoremstyle{remark} 
 
\theoremstyle{definition}
\newtheorem{example}[theorem]{Example} 
 
\newtheorem{remark}[theorem]{Remark}

\newcommand{\old}[1]{{{}}}

\graphicspath{{./Figures/}}

\def\N{\textbf{N}}

\let\eps\varepsilon 
\usepackage{xspace} 
\usepackage[margin=1in]{geometry}

\def\Thanks#1{\gdef\thefootnote{\arabic{footnote}}\thanks{#1}}
\def\ThanksComma#1{\gdef\thefootnote{\arabic{footnote},}\thanks{#1}{
}}

\newenvironment{Proof}[1]{\par\noindent{\bf Proof{#1}:}\quad}{} 

\usepackage{etoolbox}

\usepackage{xcolor}
\usepackage{pagecolor}

\title{Algorithms for Reconstructing DDoS Attack\\ Graphs Using Probabilistic Packet Marking}

\author{D. Barak-Pelleg\ThanksComma{Department of Mathematics, Ben-Gurion
University, Beer Sheva 84105, Israel.
E-mail: dinabar@post.bgu.ac.il}
\Thanks{Research supported in part by the Cyber Security Research Center, Prime Minister's Office,  and a Hillel Gauchman scholarship.}
\and
D.~Berend\ThanksComma{Departments of  Computer Science and Mathematics, Ben-Gurion
University, Beer Sheva 84105, Israel.
E-mail: berend@cs.bgu.ac.il}
\Thanks{Research supported in part by the Milken
Families Foundation Chair in Mathematics and the
Cyber Security Research Center, Prime Minister's Office.}
\and
 T.~J.~Robinson\ThanksComma{Department of Computer Science, Ben-Gurion
University, Beer Sheva 84105, Israel.
E-mail: tomjosrob@gmail.com}
\Thanks{Research supported in part by the Cyber Security Research Center, Prime Minister's office.}
\and
I.~Zimmerman\Thanks{Department of Computer Science, Ben-Gurion
University, Beer Sheva 84105, Israel.
E-mail: itamarzimm@gmail.com}}

\begin{document}

\maketitle

\begin{abstract}

DoS and DDoS attacks are widely used and pose a constant threat. Here we explore Probability Packet Marking (PPM), one of the important methods for reconstructing the  attack-graph and detect the attackers. 

We present two algorithms. Differently from others, their stopping time is not fixed a priori. It rather depends on the actual distance of the attacker from the victim.
Our first algorithm returns the graph at the earliest feasible time, and turns out to guarantee high success probability. The second algorithm enables attaining any predetermined success probability at the expense of a longer runtime.

We study the performance of the two algorithms theoretically, and compare them to other algorithms by simulation.

Finally, we consider the order in which the marks corresponding to the various edges of the attack graph are obtained by the victim. We show that, although edges closer to the victim tend to be discovered earlier in the process than farther edges, the differences are much smaller than previously thought.

	\vskip0.5em\noindent\textit{Keywords and phrases}:
DoS attack, DDoS attack, Probabilistic Packet Marking, Edge Sampling, Coupon Collector's Problem.

\end{abstract}

\setcounter{section}{0}

\section{Introduction}\label{sec:introduction}
A denial-of-service (DoS) attack is a cyber attack on a specific computer on the Internet Network. In the attack, the single attacker tries to flood the victim with fake data packets, until it collapses. Distributed denial-of-service (DDoS) expands the DoS concept by replacing a single attacker with multiple attackers. Mostly, attackers are slave machines, controlled by some trojan or malware.

These attacks are widely used, and pose a constant threat. Despite the fact that they are familiar, they still cause damage consistently. In fact, the recent invasion of Russia into Ukraine was preceded by a DDoS attack on a number of government websites and financial institutions \cite{Scroxton2022}. Among the other victims in recent years we mention Telegram \cite{EX1}, GitHub \cite{EX2}, DNS servers \cite{EX4}, Wikipedia \cite{EX5}, and various financial sector institutions
\cite{EX3}. (See also The Navy Times \cite{NavyTimes2017-1,NavyTimes2017-2}.) There is an understanding that this kind of attack is here to stay, and that it will probably increase with the rise of IOT technology \cite{IOT2,IOT,IOT3,IOT4}.
\newline
\newline
There are several defense techniques to deal with these attacks \cite{DefenseMain, Defense2020, LoukasOke, DefenseMet, DefenseMet1, DefenseMet2, DefenseMet3, DefenseMet4, DefenseMet5, DefenseMet6}. One of the main approaches is to detect the attackers on the network, and then mitigate the attack by filtering messages coming from them, or try to search the attack's origin through the slave machines.

\subsection{The Attack Graph and the IP Traceback Problem}
To find the attacker, we need to reconstruct the attack graph,
which is a tree-type graph. (In a DoS attack, the graph is just a path.) The victim is the root of the graph, the leaves are the attackers, and the internal nodes are routers on the network (connecting the attackers and the victim). The reconstruction of the attack graph is a kind of IP traceback \cite{IPTraceback,IPTraceback1,IPTraceback2}, that can be executed by numerous methods, for example, employing Markov Chains \cite{MR_CH1,MR_CH2,MR_CH3}, Deterministic Packet Marking (DPM) \cite{DPM,DPM2, DPM3, DPM4}, Probabilistic Packet Marking (PPM) \cite{BB,AKSW,PPMWMPC,LPM,EF-PPM,LWW,SS2} and others \cite{DFM}. Our research focuses on edge-sampling \cite{EdgeS1, EdgeS2}, the most common method of PPM.

\subsection{Probabilistic Packet Marking}
\label{subsection:PPM_intro}
 
Savage, Wetherall, Karlin, and Anderson \cite{AKSW}  considered several versions of PPM. Here we relate to the edge sampling version. In this version, each router along the path, upon receiving a message, and before transmitting it, randomly decides with some probability \(p < 1\), independently of all  other routers, whether to mark it or not. Each upstream router on the path either overrides the previous mark, by marking the packet itself, which happens with the same probability $p$ for all routers, or delivers the packet exactly as received, except for incrementing the distance on the previous marks which is restarted each time a router overrides the mark.  Each mark consists of an edge in the attack graph and its distance from the victim. The victim can accumulate these edges and build the attack graph. Marks may reach the victim from legitimate computers or from the attackers. Clearly, as the number of packets received from an attacker usually exceeds by far the number of packets from legitimate sources, we easily  distinguish between the two types. When a computer sends many messages to the victim, the number of markers is likely to increase, and the victim may assume that the purpose of the computer is to flood him.

\subsection{The Reconstruction Time}
\label{subsection:The Reconstruction Time}
An important question regarding the PPM algorithm is how much time (in terms of packets) is needed to reconstruct the attack graph in the case of a single attacker \cite{BarakBerend2021,SS,SS2,AKSW,LWW,ReconstFast,ReconstFast2}.
We shall refer to this variable as the {\em reconstruction time}.
 On the one hand, the longer the victim continues to collect packets, the reliability of the reconstructed path increases. On the other hand, the shorter the victim waits, the path reconstruction time decreases; finding the path sooner, the victim will be able to filter out the attacker earlier and thus reduce the overall attack power.
Savage et al.\ \cite{AKSW}   considered the runtime of the edge sampling algorithm, namely, the reconstruction time of the algorithm presented there. 
They also  bounded from above the expected  reconstruction time (for $p=1/n$) by ${\log{n}}/({p(1-p)^{n-1})}\approx en\log n$.
Saurabh and Sairam \cite{SS,SS2} suggested to stop the algorithm somewhat later than the expected reconstruction time. Namely, they suggested to add to the bound of 
Savage et al.\ a fraction of the standard deviation of the reconstruction time. They referred to the recommended stopping time in the algorithm as the Completion Condition Number.

\subsection{PPM Algorithms}  
In this work, we provide  new PPM algorithms (first presented in \cite{BB}), striving to improve other PPM algorithms  \cite{AKSW,LWW}.
Obviously, the victim cannot reconstruct the attack graph without receiving a complete path from the victim to the attacker.
We examine what happens if we reconstruct the path immediately after receiving a complete path from the victim to a certain router. In this case, the error that can occur is that the algorithm will return a subpath, which does not include the attacker. We will show that this naive approach is, perhaps somewhat surprisingly, quite good.
We later suggest a second algorithm, which  yields even better reliability,  at the expense of somewhat larger runtime.

\subsection{The Order of Received Marks}  

It seems to be taken for granted in the literature that the arrival order of routers' marks during the attack is very similar to the probabilities order (first the edge from the nearest router to the victim) \cite{SS2}. We will show that the picture is more complex, and although these orders are the most common, the probability for receiving such an order is not as dominant as is commonly thought. 

\subsection{The Structure of the Paper}

In Section~\ref{secAlgorithm} we explain in detail how PPM works and present two algorithms.  Section~\ref{secMainRes} presents theoretical results concerning our algorithms and other interesting points regarding PPM. In particular, we show that the possibility, raised in some papers, of obtaining many false, but seemingly plausible attack paths, may very seldom occur. 
Section~\ref{simulations} describes a simulation, comparing our algorithms with those presented above -- of Savage et al.\ \cite{AKSW} and of Saurabh and Sairam \cite{SS2}. The bottom line is that our algorithms enable achieving at least the same success rate as previous algorithms, but quite faster. Section~\ref{secProofs} is devoted to the proofs of the theoretical results. 
Section~\ref{technicalities} addresses several technical issues. In Section~\ref{secConclusion} we present the conclusion and future work. In Appendix~\ref{OtherSimResults} we present additional results and insights from our simulation.

\section{The Main Algorithm and a Variation}\label{secAlgorithm}
Assume first that there is a single attacker. Let $P$ be the attack graph, a path with edges $e_1, \dots, e_n$, where $e_i=\{v_{i-1}, v_i\}$ represents the edge $i$ hops from the victim.  Denote by $L(P)=n$ the length of the path.  Thus, the victim is represented by the vertex $v_0$ and the attacker by $v_n$. A~\textit{marked packet} is a packet that has been received with a mark. This mark is a triple, consisting of two consecutive vertices of~$P$, ordered in decreasing distance to  the victim, together with the hop distance. Getting a marked packet, with hop distance $i$, we know the edge $e_i$ of $P$. 
In addition to marked packets, we receive unmarked packets. We denote the ``edge'' corresponding to such packets as $e_0$. We refer to $e_i$, $1 \leq i \leq n$, as \textit{proper edges} and to $e_0$ as the \textit{dummy edge}. A~\textit{flow} of packets yields a sequence of edges received by the victim. Each element of the flow is $e_{j}, 1\le j\le n,$ with probability $p_{n-j+1}=p(1-p)^{j-1}$, and is $e_0$ with probability $(1-p)^n$.  It will sometimes convenient to use the notation  $q=1-p$. 

Throughout the paper, we assume that the length $n$ of $P$ is large and the marking probability is small. (In fact, simulations show that, for common values of $n$, the approximations obtained by taking $n\to\infty$ are pretty accurate.) Specifically, to fix ideas, we will moreover assume that $p=1/n$ (see \cite{AKSW,SS2}).

When the victim realizes he is under attack, he starts collecting the information provided by the marked packets. 
At each point in time, he has several edges of $P$, and perhaps some edges relating to packets received from legitimate users. Denote the subgraph defined by the edges he has as $G$. As the number of packets from the attacker may be assumed to exceed by far that from legitimate sources, we may assume that most edges he has belong to the path $P$. Suppose the victim has at some point a maximal subpath $P'=v_0 v_1\ldots v_i$. We assume here that $i\ge 2$; an attack path will not consist of a single edge. 
Using the tools to be explained below in Section \ref{technicalities}, he can check whether the edges he has, that do not belong to $P'$, are part of a path containing $v_0 v_1\ldots v_i$. (Those with a hop distance not exceeding $i$ may be discarded without checking.) If after removing the unrelated edges, he still has edges not belonging to $P'$,  he cannot possibly reconstruct $P$.
Thus, let a {\it full subpath} of $P$ be a subpath $P'$, starting at $v_0$, such that no other edge of the graph~$G$ belongs to a path containing~$P'$ in the network. What should the victim do when he does have such a full subpath?

In \cite{LWW}, it has been suggested to wait a while in this case. At the end of this period there are three possibilities:
\begin{itemize}
    \item No new edge has been obtained during this waiting period (or some edges have been received, which he verifies to be unrelated).
    
    In this case, we guess that the subpath we have is exactly $P$.
    
    \item The edge $\lbrace{v_i,v_{i+1}\rbrace}$ is obtained. 
    
    In this case, we again have a full subpath, which is also a candidate for being the full attack path $P$, and we start a waiting period again at this point.
    
    \item Some other edge $\lbrace{v_{j},{v_{j+1}}\rbrace}$, belonging to the same path, with $j\ge i+1$, has been obtained. 
    
    In this case we do not have a full subpath anymore, and we continue collecting edges until a full subpath is again formed.
\end{itemize}

    We mention here in passing that, in fact, it may be the case that one cannot possibly find $v_n$. The attacker can forge his IP, in which case the link $\lbrace{v_{n-1},v_n\rbrace}$ will never arrive at the victim. As done by Savage et al.\ \cite[Sec.3.1, p.298]{AKSW}, in this case we actually look for a path approximating the true attack path, namely deal with the so-called approximate traceback problem. This problem is more robust, as it does not depend on any assumptions regarding the attacker's abilities.

The waiting period discussed above is taken so as to guarantee, under a worst-case scenario, a sufficiently large probability of reconstructing the full attack  path $P$. In fact, the algorithm is easily adapted to the case of any attack graph, not necessarily a path. We simply act against each attacker as soon as he is discovered. (Note, though, that the analysis below may be not completely accurate if the attack paths of some different attackers have some shared edges.) 

As mentioned at the outset, we first deal with a single attack. Algorithm~\ref{Algorithm1} below is obtained from the algorithm of \cite {LWW} by leaving the waiting time out. Namely, if at any point, the edges we have collected form a full subpath, we guess that this subpath is the full attack path. 

\begin{algorithm}[H]
\caption{{\bf -- Basic Algorithm} \label{Algorithm1}}
\begin{algorithmic}
\STATE \textbf{Input: }{Incoming flow of packets}
\STATE \textbf{Output: }{Attack path}
\STATE $G \leftarrow \emptyset$ \;

\WHILE{$G$ \rm{does not contain a full subpath} $P'$ \\ \hspace{9.5mm}of length $\ge 2$}
\STATE Receive the next packet\;
\IF{\rm{the packet is marked by some edge} $e_i$, $i\ge 1$,}
\STATE $G \leftarrow G\cup  e_i$\; 
\ENDIF
\ENDWHILE
\STATE {return $P'$}
\end{algorithmic}
\end{algorithm}

Algorithm~\ref{Algorithm1} is optimal in the sense that it returns an answer at the earliest possible time. At no point prior to that does the victim have a feasible guess as to the attack path. However, it seems at first sight that the algorithm guarantees everything but reconstructing the correct attack path. Indeed, should we not expect seeing a short full subpath, say $v_0v_1v_2$, first, and conclude (most probably erroneously) that $v_2$ is the attacker? More generally, if at some point, before getting the whole of $P$, we have collected exactly the edges $e_1,e_2,\ldots,e_i$ for some $2\le i\le n-1$, then we err in our guess. Do all these error probabilities not add up to a very large probability?

Perhaps somewhat surprisingly, it turns out that this algorithm yields the full attack path with probability $1-o(1)$ as $n$ tends to infinity (see Theorem~\ref{th:basicacc} below).

Algorithm~\ref{Algorithm2} is a heuristic algorithm based on Algorithm~\ref{Algorithm1}. 
The accuracy level of Algorithm~\ref{Algorithm1} is fixed once the marking probability $p$ is chosen. For instance (unless the attack path is of length $2$), there is already a failure rate of at least  $2\cdot p\cdot p(1-p)$ due to the possibility that the first two arriving packets will be marked by the first two edges.  
To remedy this problem, we add, similarly to \cite{LWW}, a  waiting period after a full subpath has been obtained. This waiting time differs from that taken in \cite{LWW}, as follows.
The waiting time there is always positive. Our waiting time, on the other hand, depends on the total number of packets obtained so far. If this number is large enough, we wait no more. Thus, it is very often the case  that, once we have a full subpath, we do not wait for more packets (see the simulations in the next section). 
Only when we get to a full subpath after having received ``relatively few" packets do we indeed wait for more packets before guessing that the current subpath is $P$. We now explain how our waiting time, if any, is determined.

Suppose we have received by now $\ell$ packets and a full subpath $P'$ of length $j=L(P')$ has been generated. We decide whether or not to wait for more packets as follows. Note that, if $P'$ is not the whole of $P$, then there is (at least) another $(j+1)$-st edge in the path which  has not been obtained. The probability that none of the received   $\ell$ packet was marked by this edge  is $(1-pq^{j})^{\ell}$. We continue gathering packets until this probability is ``small enough''. Thus, let $\eps>0$ be some small parameter. We accept $P'$ as the full attack path if $(1-pq^{j})^{\ell}<\eps$. In this case, $P'$ is referred to as $\varepsilon$-\textit{full timed}.

\begin{algorithm}[H]
\caption{{\bf -- A More Accurate Heuristic Algorithm} \label{Algorithm2}}

\begin{algorithmic}
 \STATE \textbf{Input: }{Marking probability $p$,\\ \hspace{10.5mm}incoming flow of packets,\\ \hspace{10.5mm} $\varepsilon$} (small parameter controlling the accuracy)
\STATE \textbf{Output: }{Attack path}
\STATE $G \leftarrow \emptyset$ \;

\WHILE{$G$ \rm{does not contain an $\varepsilon$-full timed subpath $P'$\\ \hspace{9.5mm}of length $\ge 2$}}
\STATE Receive the next packet\;
\IF{\rm{the packet is marked by some edge} $e_i$, $i\ge 1$,}
\STATE $G \leftarrow G\cup  e_i$\; 
\ENDIF
\ENDWHILE
\STATE {return $P'$}
\end{algorithmic}
\end{algorithm}

\section{Main Results}\label{secMainRes}
Let $A$ be the event that Algorithm~\ref{Algorithm1} returns the full attack path. Let $T_1$ be the expected number of packets sent until the algorithm stops. Recall that the {\it harmonic numbers} are defined by
$$H_n=1 + 1/2 +1/3 +\cdots+ 1/n, \qquad n=1,2,3\ldots,$$
and that $H_n=\log n+O(1)$ as $n\to\infty$.

\begin{theorem}
\label{th:basicacc}
Consider Algorithm~\ref{Algorithm1}.
\begin{description}
    \item {1.}
The accuracy of the algorithm satisfies:
\begin{align*}
P(A) = 1-O\left(\frac{(\log n)^{1/(e-1)}}{n^{1/e-1/e^2}}\right)=1-o(1).
\end{align*}
\item {2.}
The expected number of packets sent until the algorithm stops is:
\begin{align*}
T_1 \leq enH_n = en\log{n}+O(n).
\end{align*}
\end{description}
\end{theorem}

\begin{remark}
\label{rem:exprun}
The bound in the second part of the theorem is an immediate corollary of the bound in \cite{AKSW}. In fact, in that paper, the same bound was obtained for a stochastically larger random variable, namely the time until all the attack path has been obtained. We bring our own proof, though, as it follows directly from the classical result of the coupon collector problem, without a need for additional calculations.
\end{remark}

 In Theorem~\ref{2succ} below, we obtain an upper bound on the average number of packets required in Algorithm~\ref{Algorithm2}, that exceeds the bound for Algorithm~\ref{Algorithm1} only by the sum of a linear term and a single global constant, depending on $\varepsilon$.
 
Let   $T_2(\varepsilon)$ be the expected number of packets sent until Algorithm~\ref{Algorithm2} stops.

\begin{theorem}
\label{2succ}
Consider Algorithm~\ref{Algorithm2}. 
\begin{align*}
T_2(\varepsilon) \le en\log{n}+\frac{1}{\varepsilon}+O(n).
\end{align*}
\end{theorem}

There is a commonly made assumption that edges arrive according to decreasing order of probability (see \cite{SS2}). However, there is a significant amount of deviation from this ordering, as Theorems~\ref{th:location} and~\ref{th:disr} below demonstrate.  If the edges appear in the order indicated by $\textbf{v}=\langle j_1, \dots, j_n \rangle$, then the {\it location} of edge $j$ is the index of $\textbf{v}$ in which it appears. Let $L_n(j)$ denote the location of edge $j$.  The \textit{normalized location} of edge $j$ is $\hat{L}_n(j)=L_n(j)/n$.

\begin{theorem}
\label{th:location}
\begin{description}
\item{1.}
The expectation $E(L_n(i))$, considered as a function of $i$, is ``anti-symmetric'' with respect to the point $\frac{n+1}{2}$, namely
\begin{align*}
    E(L_n(i))+E(L_n(n+1-i))=n+1, \qquad 1\le i\le n.
\end{align*}
\item{2.} For fixed $0\leq \beta \leq 1$ and $i=n\beta(1+o(1))$, the normalized location of edge $i$ satisfies
\begin{align*}
\lim_{n \rightarrow \infty} E(\hat{L}_n(i))=
\log\left(1+e^{\beta}\right)-\log\left(1+e^{\beta -1}\right).
\end{align*}
\end{description}
\end{theorem}

\begin{remark}
In fact, as $n\to\infty$, uniformly over $1 \leq i \leq n$:
\begin{align}
\label{average-location}
E(L_n(i))
=n\left(\log\left(1+e^{i/n}\right)-\log\left(1+e^{i/n-1}\right)\right)+O(1).
\end{align}
\end{remark}

\begin{figure}[ht]
\centering
\includegraphics[width=
4in]{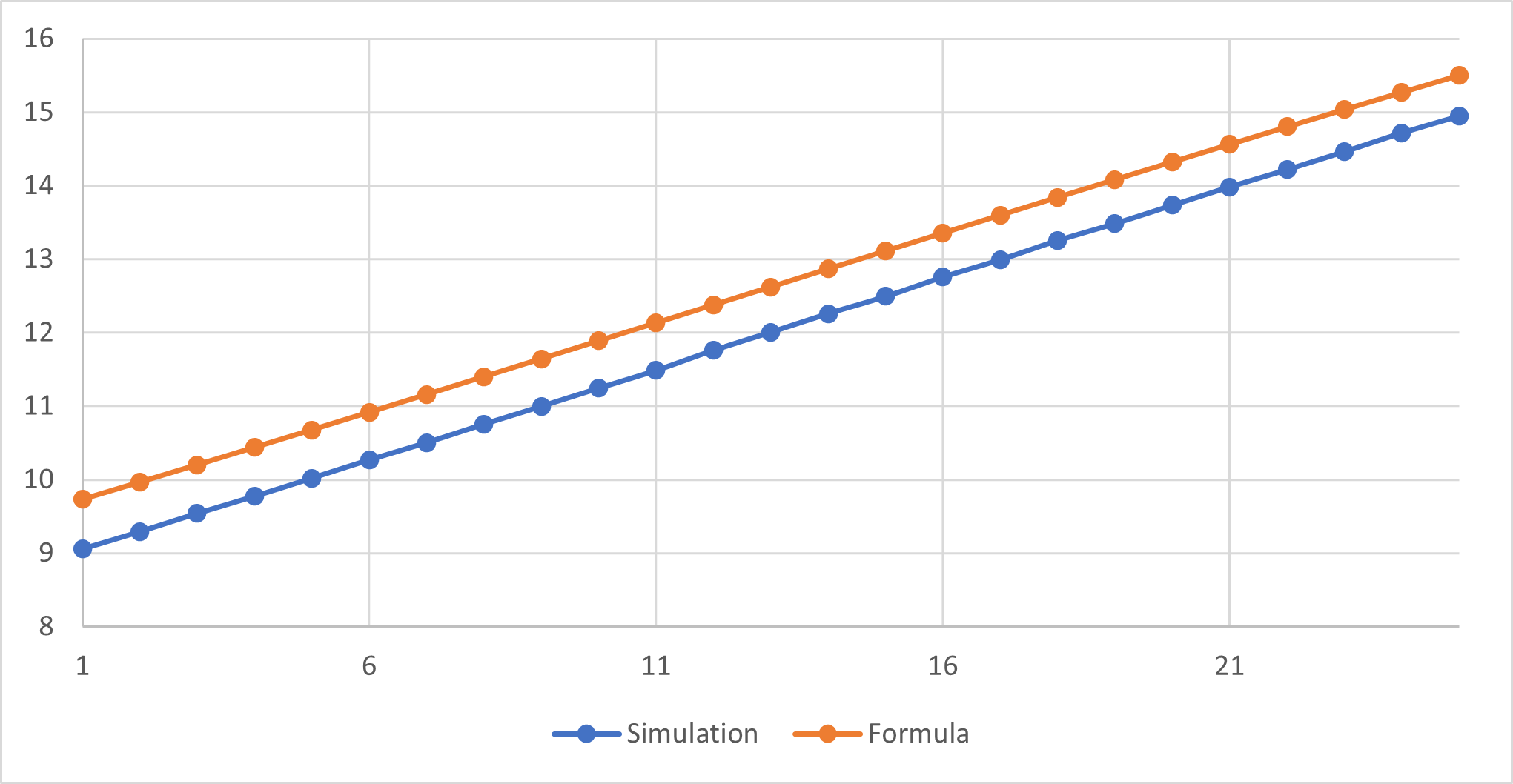}
\caption{The average location of each edge}
\label{g2.png}
\end{figure}

In Figure \ref{g2.png} we depict $E(L_n(i))$ as a function of $i$ for $n=25,  p=1/25$. The bottom graph represents the left-hand side of (\ref{average-location}), based on simulations, and the top one -- the main term on the right-hand side of~(\ref{average-location}).

\begin{example}
As $n\to\infty$, the normalized location of the least probable edge (corresponding to $\beta=1$) tends to
\begin{align*}
\log\left(\frac{e+1}{2}\right)\approx 0.62,
\end{align*}
and that of the maximal probability proper edge ($\beta=0$) tends to
\begin{align*}
\log\left(\frac{2e}{e+1}\right)
=1-\log\left(\frac{e+1}{2}\right) \approx 0.38.
\end{align*}
\end{example}

A \textit{disruption} is a pair $(i,j)$ with $1\le i < j\le n$ such that edge $i$ occurs after edge $j$.  Let $D_n(i)$ be the  number of disruptions involving $i$.  Denote by $\hat{D}_n(i)=D_n(i)/n$ the normalized number of disruptions involving $i$.

\begin{theorem}
\label{th:disr}
\begin{description}
\item{1.} The expectation $E(D_n(i))$, considered as a function of $i$, is ``symmetric'' with respect to the point $\frac{n+1}{2}$, namely  
\begin{align*}
    E(D_n(i))=E(D_n(n+1-i)) \qquad 1\le i\le n.
\end{align*}
\item{2.} For fixed $0\leq \beta \leq 1$ and $i=n\beta(1+o(1))$, the normalized expected number of disruptions involving edge $i$ satisfies:
\begin{align*}
\lim_{n \rightarrow \infty} E(\hat{D}_n(i))&=
1+\log{4}-\log\left(1+e^{1-\beta}\right)-\log\left(1+e^\beta\right).
\end{align*}
\end{description}
\end{theorem}

\begin{remark}
In the proof we show that, moreover, as ${n\to\infty}$, 
uniformly over $1 \leq i \leq n$:
\begin{equation}
    \label{number of deviations for edge i}
    \begin{aligned}
E(D_{n}(i))=&\:n\left(
1+\log 4 -\log{\left(1+e^{1-i/n}\right)}-\log{\left(1+e^{i/n}\right)}\right)+O(1).
\end{aligned}
\end{equation}
\end{remark}

\begin{figure}[ht]
\centering
\includegraphics[width=4in]{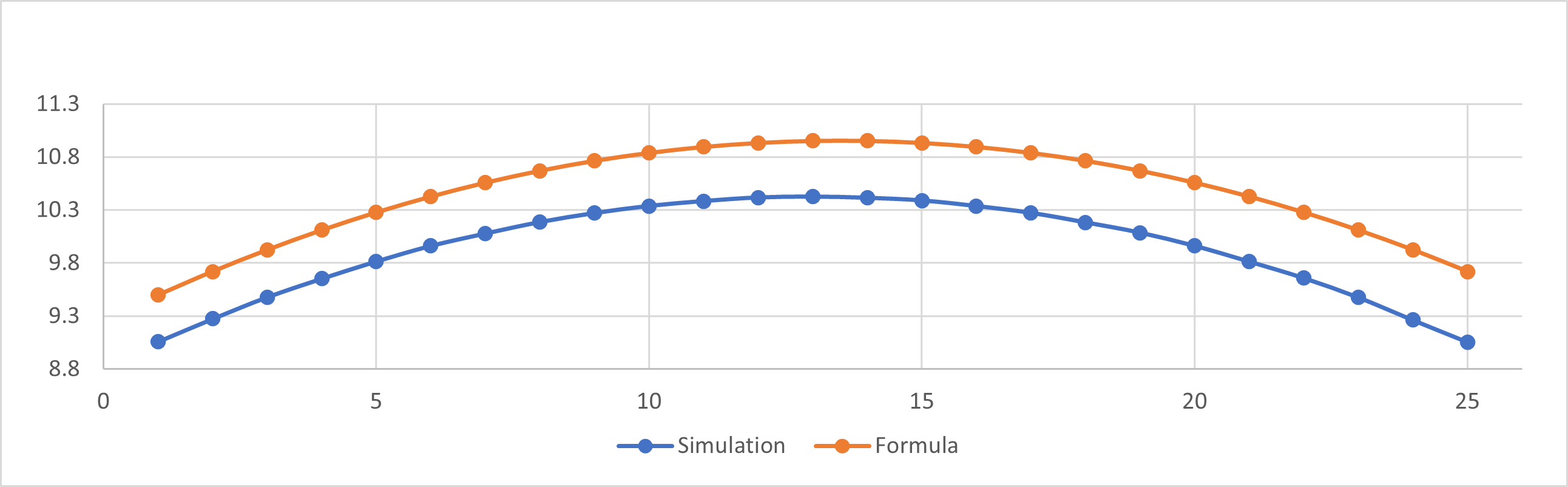}
\caption{The average number of edges obtained at the ``wrong" location with respect to edge~$i$}
\label{g3.png}
\end{figure}

In Figure \ref{g3.png} we depict $E(D_n(i))$ as a function of $i$ for $n=25, p=1/25$. The top graph is the main term on the right-hand side of (\ref{number of deviations for edge i}) and the bottom one is based on simulations.

\begin{remark}
In the limit as $n\to\infty$, the expected normalized disruption is unimodal in $\beta$, and is maximized at $\beta=1/2$, and minimized at $\beta=0,1$.
\end{remark}

Let $B(i_1, \dots, i_n)$ denote the event whereby the first edge to be received is $i_1$, the second is $i_2$ and so forth.  Let $\overline{B}=B(1,2,\dots,n)$ denote the event in which the edges are received in descending order of probability, and $\underline{B}=B(n,n-1, \dots, 1)$ is the event in which the edges are received in increasing order of probability.  

We recall the dilogarithm function $\text{Li}_2$, defined by
$$\text{Li}_2(z)=-\int\limits_0^z \frac{\log(1-t)}t dt,\qquad z\in\bf{C}.$$
For $|z|\le 1$, which is where we will need the function, we have an alternative expression:
$$\text{Li}_2(z)=\sum\limits_{j=0}^{\infty} \frac{z^j}{j^2}$$
(see, for example, \cite[Sec A.2.1.(1)]{Lewin}).

\begin{theorem}
\label{maxB}
\begin{description}
\item{1. }For any ordering $B$
$$P(\underline{B}) \leq P(B) \leq P(\overline{B}).$$
\item{2. }As $n\to\infty$,
\begin{align*}
    P(\overline{B})&=\Theta\left(e^{\left({\pi^2}/{6}
-\mathrm{Li}_2\left(e^{-1}\right)\right)n}/n^{n-1/2}\right)\\
&\simeq \Theta\left({e^{1.236n}}/{n^n}\right).
\end{align*}
\item{3. }As $n\to\infty$,
\begin{align*}
    P(\underline{B})&=\Theta\left({e^{\left({\pi^2}/{6}-\mathrm{Li}_2\left(e^{-1}\right)-{1}/{2}\right)n}}/{n^{n-1/2}}\right)\\
&\simeq \Theta\left({e^{0.736n}}/{n^n}\right).
\end{align*}
\item{4.}
$\displaystyle{ \:\:P(\underline{B})/P(\overline{B})=q^{n(n-1)/2}=e^{-n/2+1/4}\left(1+o(1)\right).}$
\end{description}
\end{theorem}

\begin{remark}
The average of the probabilities over all possible orderings is ${1}/{n!}$, which by Stirling's approximation is (up to a $C\sqrt{n}$ factor) $e^n/n^n$. By the theorem, the probabilities actually fluctuate between about $e^{0.236n}$ times this average at the top and $e^{-0.264n}$ at the bottom. Namely:
\begin{align*}
    e^{-0.264n}\cdot\frac{1}{n!}\lesssim P(B) \lesssim e^{0.236n}\cdot\frac{1}{n!}.
\end{align*} 
Thus, while there is an exponentially large range of variation of these probabilities, this range is small relative to $n!$.
\end{remark}

\section{Simulation Results}\label{simulations}
We have performed a simulation of the process of collecting the  markings in a DoS attack in order to compare Algorithms~\ref{Algorithm1} and \ref{Algorithm2} with those of Savage et al.\ \cite{AKSW} and Saurabh and Sairam \cite{SS2}. 

Let us first recall the two latter algorithms. Savage et al.\ \cite{AKSW} have not explicitly stated an algorithm. 
As mentioned in the introduction, they showed that the expected reconstruction time is bounded above  (for $p=1/n$) by ${\log{n}}/({p(1-p)^{n-1})}\approx en\log n$. This may be interpreted, and has been indeed interpreted by some (\cite{SS2,Kiremire2014}), to mean that the victim should try to reconstruct the attack path after receiving this number of packets. We will also refer to it as the algorithm suggested by Savage et al. We denote it by SWKA. Saurabh and Sairam suggested to attempt the reconstruction after receiving
\begin{equation*}
    \frac{\log n}{p(1-p)^{n-1}}+\frac13\cdot \sqrt{\sum_{i=1}^n\frac{1-\sum_{j=1}^ip_j}{\left(\sum_{j=1}^ip_j\right)^2}}.
\end{equation*}
packets. We refer to this algorithm as S\&S. As for Algorithm~\ref{Algorithm2}, we have tested it for  $\eps=0.1, 0.05$.

\begin{figure}[ht]
\centering
\includegraphics[width=3.1in]{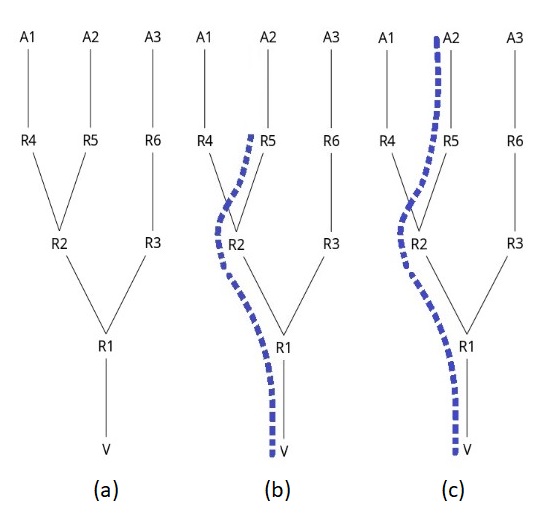}
\caption{The edges obtained by time: (a) $t_1=0$, (b) $t_1=13$, (c) $t_2=16$.}
\label{toy}
\end{figure}

Before describing the simulation, we will present a toy example. Consider the network in Figure~\ref{toy}. We focus on the attack by $A_2$. We have $n=4$, and take accordingly $p=1/4$. At time $t_1=0$, the attack starts (Figure~\ref{toy}(a)). Suppose that we obtain edges during the run as follows:\\
$$e_2, e_3, e_0, e_2, 
e_1, e_2, e_3, e_0, 
e_2, e_2, e_1, e_3, 
e_0, e_0, e_4, e_1, e_2.$$
According to Savage et al., we try to reconstruct the attack path at time
$$t_1=\frac{\log 4}{1/4\cdot(1-1/4)^3}\approx 13$$
(see Figure~\ref{toy}(b)). By this time, we have obtained the edges $e_1, e_2,$ and $e_3$, but not $e_4$. Thus, SWKA returns erroneously the path $vR_1R_2R_5$ as the supposed attack path and $R_5$ as the corresponding supposed attacker. S\&S waits until

\begin{align*}
t_2&=t_1+\frac13\cdot \sqrt{\sum_{i=1}^4\frac{1-\sum_{j=1}^i1/4\cdot(1-1/4)^{4-j}}{\left(\sum_{j=1}^i1/4\cdot(1-1/4)^{4-j}\right)^2}}\approx 16
\end{align*}
and decides correctly that $A_2$ is the attacker (see Figure~\ref{toy}(c)). Algorithm~\ref{Algorithm1} returns the attack graph at time $t_3=15$ and obtains the full attack path. Altogether, in this example, SWKA returns an incorrect result, while Algorithm~\ref{Algorithm1} and S\&S return the correct path, with Algorithm~\ref{Algorithm1} finishing slightly earlier.

In our experiment, the number of coupons is $n=25$, the marking probability is $p=1/25$, and the number of iterations  is $N=10^7$. In each iteration, we receive packets arriving along the path $P$ until we get the full attack path~$P$. 

At the end of each iteration we have recorded the number of packets  received until we have obtained all the edges of the attack graph. For each of the five algorithms (counting Algorithm~\ref{Algorithm2} once for every $\eps$), we have recorded the number of received packets   and whether or not the algorithm has been successful in obtaining $P$. 
During the process we also record the following data:  
(i) Each time the edges we have so far constitute a full subpath of length $\ge 2$, we have recorded its length. In particular, we have recorded  the length of the first full subpath as that returned by Algorithm~\ref{Algorithm1}. 
(ii) At the end of each iteration we have recorded the number of times we have encountered a full subpaths during the process. 
(iii)  Each time  the graph  returned by Algorithms  SWKA or S\&S was  not  $P$ we have recorded whether of not the graph was a full subpath.
 
For example, suppose that the order in which the edges have arrived (suppressing repeated arrivals) is:
\begin{align*}
    &6, 14, 5, 3, 12, 8, 1, 4, 11, 7, 9, 13, 10, {\bf{2}},\\
    &16, 19, 23, 17, 18, 21, 22, 20, {\bf{15}},{\bf{24}},{\bf{25}}.
\end{align*}

The first time we had a full subpath in this example is after receiving edge $2$, and the length of this subpath is $14$, the second time -- after receiving edge $15$, with corresponding length $23$, the third -- $24$, with length $24$, and the fourth -- $25$, with length $25$.
In this case, Algorithm~\ref{Algorithm1} would return a full subpath of length $14$. In such a case we record that the full subpaths encountered during the process are of lengths $14$, $23$, $24$ and $25$. We also record that we had four full subpaths during this iteration.
 
The main results of the simulation are presented in
Table~\ref{Table1}. We first list, for Algorithms \ref{Algorithm1} and \ref{Algorithm2}, the average number of packets received until each algorithm finished. For SWKA and S\&S, we record the predetermined fixed numbers of packets received until the algorithms terminate. Then we present the success rate of each algorithm.

We have given SWKA and S\&S a big head start in our experiment. Both these algorithms use the number $n$ when calculating their stopping time. Of course, this number is unknown in advance, so some bound or estimate has to be used. If this estimate is too large, one may expect these algorithms to stop very late, while if it is too small -- to stop too early. Our algorithms do not depend on information about $n$. In the experiment, we let SWKA and S\&S know the exact value of $n$.

 \begin{table}[ht]
\renewcommand{\arraystretch}{1.3}
\centering
\begin{tabular}{|p{2cm}|l||c|c|}
\hline
\multicolumn{2}{|c||}{Algorithms}&
\multicolumn{1}{c|}{\makecell{Average Number of\\Received Packets} }&
\multicolumn{1}{c|}{Success Rate}\\
\hline
\multicolumn{2}{|c||}{Algorithm \ref{Algorithm1}}&
\multicolumn{1}{c|}{$167$}&
\multicolumn{1}{c|}{$0.87$}\\
 \hline
\multirow{2}{=}
{\parbox{5cm}{Algorithm~2}}
&$\varepsilon=0.1$ & $191$ & $0.97$ \\ \cline{2-4}
&$\varepsilon=0.05$ & $222$ &$0.99$ \\ 
\hline
\multicolumn{2}{|c||}{SWKA \cite{AKSW}}&
\multicolumn{1}{c|}{$214$}&
\multicolumn{1}{c|}{$0.77$}\\
\hline
\multicolumn{2}{|c||}{S\&S \cite{SS2}}&
\multicolumn{1}{c|}{$241$}&
\multicolumn{1}{c|}{$0.86$}\\
 \hline
\end{tabular}\label{Table1}
\caption{Performance of the algorithms for $n=25, p=1/25$. Average number of received packets until  collecting all edges is $178$.}
\end{table}

We observe that the success rate of Algorithm~\ref{Algorithm1} is higher than that of SWKA \cite{AKSW} and is at least as high as that of S\&S \cite{SS2} (even though the last two have a head start, as explained above).
 This is achieved with a much smaller number of packets we need to wait for. As for Algorithm~\ref{Algorithm2}, in its more conservative version with $\eps=0.05$, it requires time similar to that of SWKA and S\&S, but attains a success rate close to 100\%.

\begin{figure*}[t!]
\centering
\includegraphics[width=3.1in]{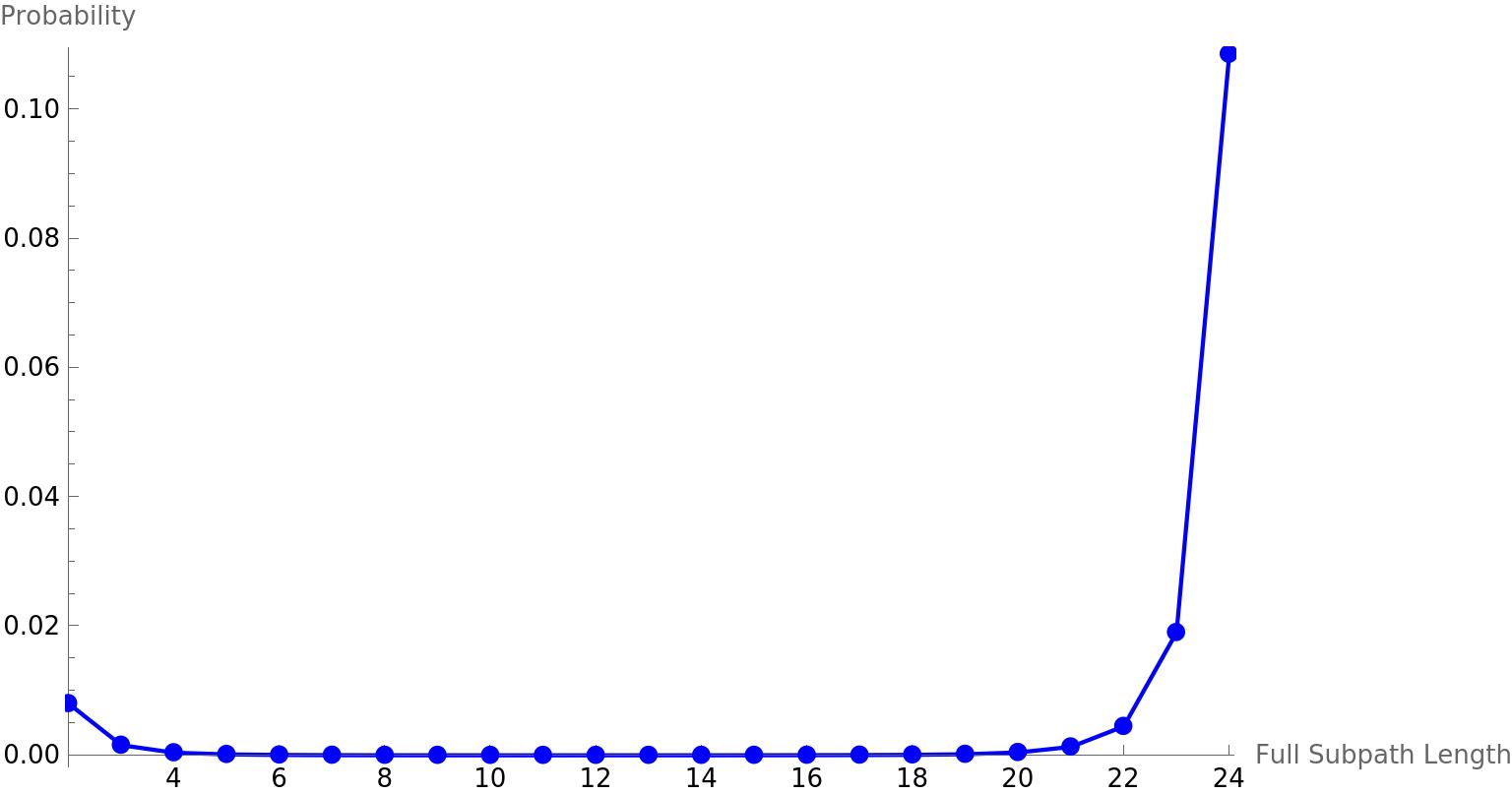}%
\label{Fig1}
\hfil
\includegraphics[width=3.1in]{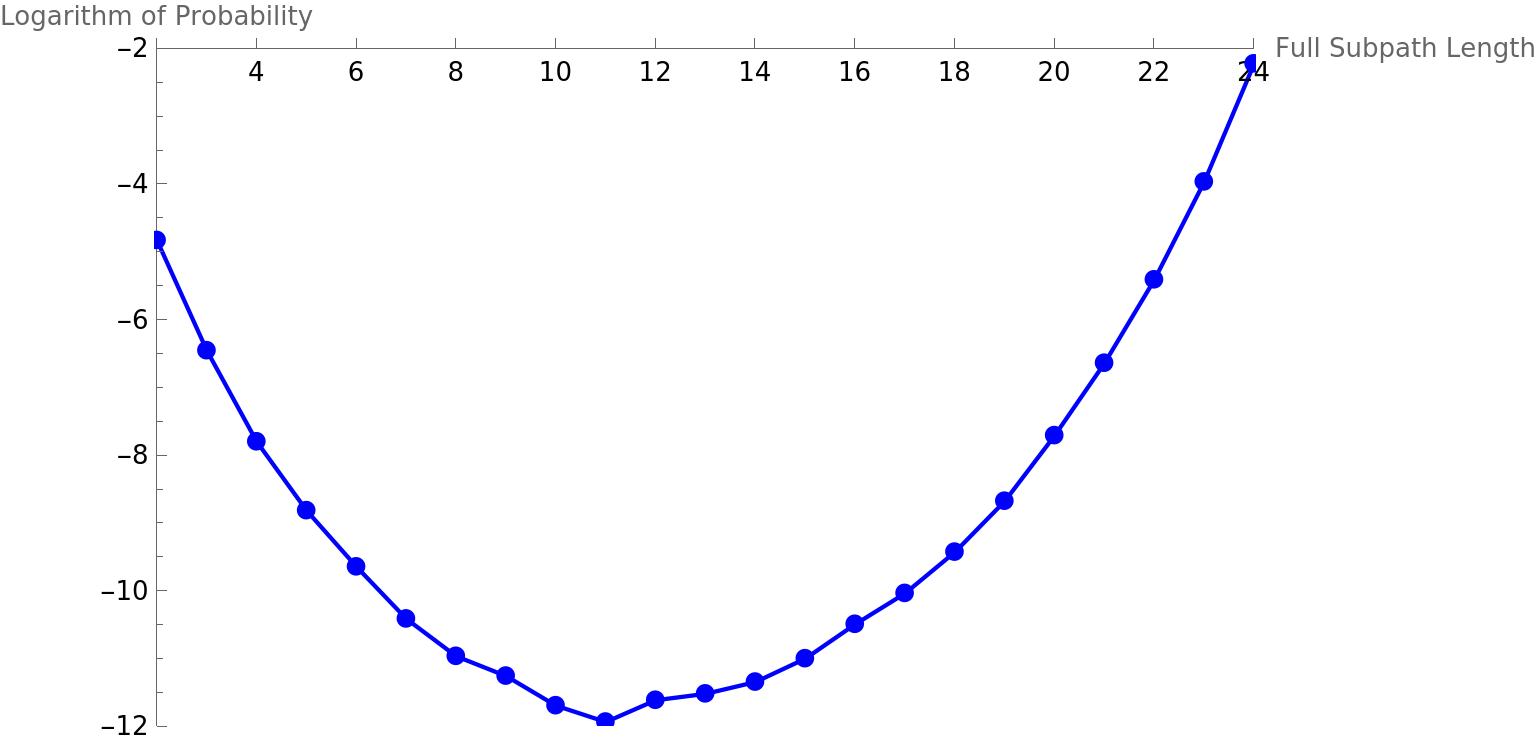}%
\label{logFig1}
\caption{Probability of obtaining distracting full subpaths of  lengths $2$ to $24$ for $n=25$ and $p=0.4$.}
\label{fig_sim}
\end{figure*}

In Figure \ref{Fig1} we depict the probabilities of getting full subpaths of all lengths up to $n-1$.  Note that these events are not pairwise disjoint; in principle, one may obtain a full subpath several times during the process. (Obviously, waiting enough, we will always get the full $P$ with probability~$1$, so that we exclude this point from the graph.)  As can be seen, it is relatively common to get the full subpaths of length  $24$, the lengths $2$ and $23$ also may appear from time to time, but all lengths in the range $3$ to $22$ have negligible probabilities. 

There is a vast difference between the size magnitude of the probabilities presented in  Figure~\ref{Fig1}. To  better view these probabilities,  we switch to a semi-logarithmic plot in Figure~\ref{logFig1}. 
For example, the probability of having at some point a full subpath of length~$6$ is $6.5\cdot 10^{-5}$, and its logarithm is $-9.6$. Observe that the least likely full subpath length is $11$, and the probability of obtaining such a subpath is about $e^{-11.9}=6.6\cdot 10^{-6}$. Moreover, in lengths 3 to 22, the maximum is about $e^{-5.4}=4.5\cdot 10^{-3}$. See Table~\ref{table4} in Appendix~\ref{OtherSimResults} for more detail.

In \cite{LWW}, the possible scenarios in which $P$ is obtained are described as follows: 
(i) The best case scenario, in which the first time we have a full subpath we get $P$. 
(ii) The worst case scenario, in which during the reconstruction of the attack graph, we always get a full subpath. Namely, we first get a full subpath of length $2$, then of $3$, and so on, until finally we obtain $P$. 
(iii) The last case scenario is the average case scenario, comprising all the other possibilities of arriving at $P$.

To check the probability of each of these events, we found the number of times  a full subpath was obtained in the course of each iteration. We present our results in Table~\ref{table2}. In the first column we present the number of full subpaths one may obtain during the reconstruction of $P$. In the second column we present the probability of obtaining that many full subpaths during the reconstruction of the graph.

 \begin{table}[ht]
\renewcommand{\arraystretch}{1.3}
\label{table2}
\centering
\begin{tabular}{|c||c|}
\hline
 \makecell{Number of Full Subpaths\\Obtained in an Iteration}&Probability \\
\hline
$1$ & $0.87$  \\ \hline
$2$ & $0.12$  \\ \hline
$3$ & $0.01$  \\ \hline
$4$ & $10^{-3}$  \\ \hline
$5$ & $10^{-4}$  \\ \hline
$6$ & $10^{-5}$  \\ \hline
$7$ & $10^{-6}$  \\ \hline
$8$ to $24$ & $0$  \\ \hline
\end{tabular}
\caption{Number of full subpaths vs.  probability of obtaining that many during the reconstruction of $P$.}
\end{table}

We observe that the best case scenario is by far the most common, the average case scenario being roughly six times less likely, and the worst case scenario having a negligible probability. This observation is corroborated by our theoretical results in the next section.

It is interesting to compare our results with relevant theoretical results. In \cite{BarakBerend2021} we have calculated the asymptotic distribution of the reconstruction time as $n$ tends to infinity. By \cite[Thm.3.b]{BarakBerend2021}, the main term of the expectation for $p=1/n$ is $$en(\log n-
\log\log n +\gamma).$$
For $n=25$ this main term equals $179$, which  is very close to our simulation sample mean, which is $178$.

\section{Proofs}\label{secProofs}

\begin{lemma}
\label{prop:bin}
If $1 \le b\le \log j$ then 
\begin{align*}
\binom{b+j}{j} \geq j^{b-\log{b}}.
\end{align*}
\end{lemma}

\begin{Proof}{} 
Note that
\begin{align*}
\binom{b+j}{j}&=\binom{b+j}{b}
=\frac{(b+j)(b-1+j)\cdots(1+j)}{b!}
\geq \left(\frac{j}{b}\right)^{b}
=\frac{j^b}{e^{b\log b}}
\geq \frac{j^{b}}{j^{\log b}}.
\end{align*}

\qed
\end{Proof}

Consider a ``classical'' coupon collector, namely the case collecting $n$ coupons, each coupon collected with the same probability~$1/n$.
Suppose the coupons are labeled by the numbers $1,2,\ldots,n$, and we are interested in the time until coupons $1,2,\ldots,j$ have all been collected. Denote this time by $S(j)$. Let $Y_i, 1 \leq i \leq j$, be the number of steps needed to receive the $i$-th distinct coupon (among the first $j$) after $(i-1)$ distinct coupons have been received. $Y_i$ is distributed geometrically with parameter $\lambda_i=(j-i+1)/n$. The variables $Y_i$ are independent. Clearly, $S(j)=Y_1+Y_2+\cdots+Y_{j}$.   
\begin{proposition}
\label{ch}
For constant $c\leq 1$, $b\geq 1$:
\begin{align*}
P(S(j) \leq cn\log j) &\leq {j^{cb}}/{\tbinom{b+j}{j}}, \qquad j=1,2,\dots,n.
\end{align*}
\end{proposition}
\begin{Proof}{} 
According to Chernoff's inequality, specialized to the case of a geometric random variable \cite{C,Z}:
\begin{equation}\label{Y-label 1}
\begin{aligned}
P(S(j)\leq a) 
&\leq \underset{t > 0}{\text{min}}\,  e^{ta}\prod_{i=1}^{j}E\left[e^{-tY_i}\right]
= \underset{t > 0}{\text{min}}\, e^{ta}\prod_{i=1}^{j}
\frac{\lambda_ie^{-t}}{1-(1-\lambda_i)e^{-t}}.
\end{aligned}
\end{equation}
For $1\le i\le n$, letting $t=b\log\left(1+\frac{1}{n}\right)$,  and using Bernoulli's inequality, 
we obtain:
\begin{equation}\label{Y-label 2}
\begin{aligned}
\frac{\lambda_ie^{-t}}{1-(1-\lambda_i)e^{-t}}
&= \frac{\lambda_i\left(1+\frac{1}{n}\right)^{-b}}{1-(1-\lambda_i)\left(1+\frac{1}{n}\right)^{-b}}
= \frac{\left(j-i+1\right)/\left(n\right)}{\left(1+\frac{1}{n}\right)^{b}-\left(1-\frac{j-i+1}{n}\right)}\\
&= \frac{j-i+1}
{n\left(1+\frac{1}{n}\right)^{b}-n+j-i+1} 
\leq \frac{j-i+1}{b+j-i+1}.
\end{aligned}
\end{equation}
Therefore, by (\ref{Y-label 1}) and (\ref{Y-label 2})
\begin{align*}
P(S(j) \leq a) &\leq  e^{ta}\prod_{i=1}^{j}\frac{\lambda_ie^{-t}}{1-(1-\lambda_i)e^{-t}}
\leq e^{ta}\prod_{i=1}^{j}
\frac{j-i+1}{b+j-i+1} \\
&= e^{ta}/{\tbinom{b+j}{j}}
= \left(1+{1}/{n}\right)^{ab}/{\tbinom{b+j}{j}}.
\end{align*}
Finally, letting $a=cn\log j $:
$$
P(S(j) \leq cn\log j) \leq {j^{cb}}/{\tbinom{b+j}{j}}.
$$\qed
\end{Proof}

\newpage

\begin{Proof}{ of Theorem~\ref{th:basicacc}}
\begin{description}
\item{1. }Let $A'$ be the event that Algorithm~\ref{Algorithm1} fails to return the full path.  Let $F_j$, $0\le i\le n-2$, be the event that the $(j+1)$-st edge is the first edge received.  Then   
\begin{align*}
P(A') =\sum_{j=0}^{n-2} P(A'|F_j)P(F_j).
\end{align*}
For $0 \leq c=c(n) \leq 1$, to be determined later (see (\ref{f}) and (\ref{c}) below), let $\alpha=cn\log{j}$.
Let $B_j$
be the event that that edges $1, \dots, j$ are collected in at most $\alpha$ steps.
Now, to bound the probability of $B_j$, let us compare the given  process of coupon collecting to the ``classical" process. 
Denote by $T_j$ the time until we obtain the coupons indexed $1,\ldots ,j$ in the classical process, and by $T'_j$, the same time in our (the given) process. Clearly,
 for every $\alpha> 0$, we have $P(T_j >\alpha)\le P(T'_j >\alpha)$. Therefore, $T'_j$ is stochastically larger than $T_j$ \cite[p. 89]{BSS}.
  Thus, 
 \begin{align*}
P(B_j) \leq P(S(j) \leq \alpha) \leq {j^{cb}}/{\tbinom{b+j}{j}},
\end{align*}
where the right inequality is obtained by using Proposition~\ref{ch}, with
\begin{align}
\label{f}
b=b(n)=(1-1/e)\log{n}.
\end{align}  
Let $j \geq n^{1-1/e}$. We have $\log{j} \geq b$, and by Lemma~\ref{prop:bin} 
\begin{align}
\label{B}
P(B_j) \leq j^{cb-b+\log{b}}.
\end{align}

Further, let $C_j$ be the event that the least probable edge was collected in $\leq \alpha$ steps. Then 
\begin{equation}\label{Cj}
\begin{aligned}
P(C_j ) &=1-(1-pq^{n-1})^\alpha 
\geq 1-\left(1-\frac{1}{ne}\right)^\alpha
= 1-\left(\left(1-\frac{1}{ne}\right)^{en}\right)^{{c\log{j}}/{e}}\\ 
&\geq 1-\left(\frac1e\right)^{{c\log{j}}/{e}} 
 = 1-\frac{1}{j^{{c}/{e}}}.
\end{aligned}
\end{equation}
Suppose that, under $F_j$, we consider the event $A'$. For this event to occur, we must have at some point exactly all edges $e_1, e_2,\ldots, e_i$ for some $j+1\le i\le n-1$. If this happened after at most $\alpha$ steps, then after $\alpha$ steps we have certainly obtained all edges $e_1, e_2,\ldots, e_j$, so that $B_j$ has occurred. If this happened after $\alpha$ steps or more, then after $\alpha$ we certainly have not obtained $e_n$,  so that $C_j$ has not occurred. It follows that, under $F_j$, we have $A'\subseteq \overline{C_j} \cup B_j$. By the union bound, and since $j \geq n^{1-1/e}$, we obtain from (\ref{B}) and (\ref{Cj}):   
\begin{align*}
P(A'|F_j) &\leq 1-P(C_j)+P(B_j)
 \leq \frac{1}{j^{{c}/{e}}}+j^{cb-b+\log{b}}.
\end{align*}
We equate the powers in the two terms, namely we choose
\begin{align}
\label{c}
c=
\frac{b-\log{b}
}
{b+1/e},
\end{align}
so that
\begin{align}\label{thm3.1Label8.5}
P(A'|F_j)&\leq {2}/j^{\tfrac{b-\log{b}}{e b+1}}.
\end{align}
Therefore:
\begin{equation}\label{th:basicacc-label-A-1}
\begin{aligned}
P(A') =&\:\sum_{j=0}^{n-2} P(A'|F_j)P(F_j)
= \sum_{j=0}^{n^{1-1/e}-1} P(A'|F_j)P(F_j)
+\sum_{j=n^{1-1/e}}^{n-2} P(A'|F_j)P(F_j)\\
\leq& \sum_{j=0}^{n^{1-1/e}-1} pq^{j}
+\sum_{j=n^{1-1/e}}^{n-2} P(A'|F_j)P(F_j)\\
=&\: 1-\left(1-1/n\right)^{n^{1-1/e}} 
+\sum_{j=n^{1-1/e}}^{n-2} P(A'|F_j)P(F_j).
\end{aligned}
\end{equation}
Consider the second addend on the right-hand side of (\ref{th:basicacc-label-A-1}). For  $n\ge 2$
\begin{equation}\label{th:basicacc-label-A-2}
    \left(\left(1-\frac{1}{n}\right)^n\right)^{n^{-1/\varepsilon}}
    \ge \left(e^{-2}\right)^{n^{-1/\varepsilon}}
    \ge 1-2{n^{-1/\varepsilon}}.
\end{equation}
A routine calculation shows that for sufficiently large $b$ (and therefore for a sufficiently large $n$)
\begin{align}
\label{wc}
    \frac{1}{e}-\frac{1}{e^2}-\frac{\log{b}}{eb+1}
    \leq \left(1-\frac{1}{e}\right)
    \left(\frac{b-\log{b}}{eb+1}\right).
\end{align}
Thus, by (\ref{f}) and (\ref{thm3.1Label8.5})-(\ref{wc})
\begin{equation*}
    \begin{aligned}
P(A') &=O\left( {n^{-1/e}}+\frac{1}{n}\sum_{j=n^{1-1/e}}^{n}{1}/j^{\tfrac{b-\log{b}}{eb+1}}\right)
=O\left({1/n^{1/e}}+{1}/{n^{(1-1/e)\tfrac{b-\log{b}}{eb+1}}}\right)\\
&=O\left({1/n^{1/e}}+{1/n^{{1}/{e}-{1}/{e^2}-\tfrac{\log{b}}{eb+1}}}\right)
=O\left(\frac{(\log{n})^{1/(e-1)}}{n^{1/e-1/e^2}}\right) 
.
\end{aligned}
\end{equation*}
\item{2. }
Will follow the same line of proof as that at the beginning of the proof of the preceding part. We compare our  process of coupon collecting to the following new process of collecting $n$ coupons: At each step we either receive  one of the coupons with equal probability of $1/(en)$, or, with probability $1-1/e$,  receive no coupon. 
Denote by $T_n^"$ the time at which we have finished collecting all coupons in the new process.
Clearly, $T_n^"$ is stochastically larger than  $T'_n$,  and therefore the expectation in our process is at most that in the new process, namely $enH_n$. This completes the proof.

\end{description}

\qed
\end{Proof}

Consider collecting $n$ coupons, where each coupon has the same probability  $p'\le 1/n$ of being selected at each round.  Let $T$ be the number of rounds until all coupons have been collected.  Let 
\begin{align*}
T'= \begin{cases} 
      T-E(T), & \qquad  T\geq E(T), \\
      0, &\qquad \rm{otherwise}. 
      \end{cases}
\end{align*}

\begin{proposition}
\label{tee2}
$E(T') =O(1/p').$

\end{proposition}

\begin{Proof}{}
Let $A_i$, $0 \leq i \leq n$, be the event whereby exactly $n-i$ coupons  have been  collected in the first $\alpha=\lceil E(T) \rceil$ rounds.  Then using the fact: $\binom{n}{i} \leq \left(\frac{en}{i}\right)^i$, which may be easily derived using Stirling, we get
\begin{align*}
    E(T') &=\sum_{i=0}^{n}P(A_i)E(T'\vert A_i)
    \le \sum_{i=1}^{n}P(A_i)\cdot (H_i/p'+1)\\
    &\leq 1+\frac{1}{p'}\sum_{i=1}^{n}H_i\binom{n}{i}\left(1-ip'\right)^{\alpha}
    \leq 1+\frac{1}{p'}\sum_{i=1}^{n}H_i\left(\frac{en}{i}\right)^i\left(1-ip'\right)^{\alpha}\\
    &\leq 1+\frac{1}{p'}\sum_{i=1}^{n}H_i\left(\frac{en}{i}\right)^i\left(1-ip' \right)^{{\log n}/{p'}}
    \leq 1+\frac{1}{p'}\sum_{i=1}^{n}H_i\left(\frac{en}{i}\right)^i n^{-i}\\
    &\leq 1+\frac{1}{p'}\sum_{i=1}^{n}H_i\left(\frac{e}{i}\right)^i 
    =O(1/p').
\end{align*}

\qed
\end{Proof}

Fix $\varepsilon >0$. Let

\begin{align*}
    w(j,\varepsilon)=\min\left\{\ell \in \N:  \left(1-pq^{j}\right)^{\ell} \leq \varepsilon \right\},
\end{align*}

or, equivalently
\begin{equation}\label{w(n,epsil)}
w(j,\varepsilon)=\left\lceil\log\eps / \log\left(1-pq^j\right)\right\rceil.
\end{equation}

\begin{Proof}{ of Theorem~\ref{2succ}}
We have:
\begin{align*}
    w(n,\eps)\leq \frac{-\log{\eps}}{pq^n}+1.
\end{align*}
Since, for any $a>0$, the expression $\frac{-\log{x}}{a}-\frac{1}{x}$ is maximized at $x=a$, we have:
\begin{align*}
    w(n,\eps) 
    &\leq\frac{1}{\eps}-\frac{\log{(pq^n)}}{pq^n}-\frac{1}{pq^n}+1\\
    &=\frac{1}{\eps}+(en+O(1))\cdot(\log(en+O(1))-1)+1\\
    &=\frac{1}{\eps}+en\log n +O(\log n).
\end{align*}

Consider the following auxiliary coupon collecting process. Given are $n$ coupons with equal probability $p'$ of being collected at each stage. Let $M'=\max\left(w(n,\varepsilon),H_n/p'\right)$. The process we consider is to first collect $M'$ coupons, even if we have finished collecting all $n$ types earlier. Next, we continue collecting further coupons if necessary until all types have been collected. Denote the expected running time of this process by $T_3(\varepsilon)$.  Letting $T'$ be as in Proposition~\ref{tee2}, with $p'=pq^{n-1}$, we have
\begin{align*}
T_2(\varepsilon) \leq T_3(\varepsilon) \le M'+ E(T')\le\frac{1}{\varepsilon}+en\log{n}+O(n).
\end{align*}

This completes the proof.

\qed
\end{Proof}

\begin{lemma}
\label{lem:loc}
For $1 \leq i \leq n$: 
\begin{align}
\label{Int1}
\abs{
\sum_{j=1}^{i-1}
\frac{1}{1+(1-\tfrac{1}{n})^{j-i}}
-\int_{0}^{\tfrac{i-1}{n}}\frac{n}{1+(1-\tfrac{1}{n})^{ny-i}}dy} \leq \frac{1}{2},
\end{align}
\begin{align}
\label{Int2}
\abs{
\sum_{j =i}^{n}
\frac{1}{1+(1-\tfrac{1}{n})^{i-j}}
-\int_{\tfrac{i}{n}}^{1}\frac{n}{1+(1-\tfrac{1}{n})^{i-ny}}dy} \leq \frac{1}{2},
\end{align}
and
\begin{align}
\label{Int3}
\abs{
\sum_{j \neq i}
\frac{1}{1+(1-\tfrac{1}{n})^{j-i}}
-\int_{0}^{1}\frac{n}{1+(1-\tfrac{1}{n})^{ny-i}}dy} \leq \frac{1}{2}.
\end{align}
\end{lemma}
\begin{Proof}{} 
Denote $f(x)=1/(1+(1-1/n)^{x-i})$. 
As $f$ is increasing and positive
\begin{align*}
0 \leq \sum_{j = 1}^{i-1}
f(j)
-\int_{0}^{i-1}f(x)dx
&\leq \sum_{j=1}^{i-1}\left(f(j)-f(j-1)\right)\\
&\le1\cdot f(i-1) \leq 1/2.
\end{align*}
Substituting $x=ny$, we obtain
\begin{align*}
\int_{0}^{i-1}\frac{1}{1+(1-\tfrac{1}{n})^{x-i}}dx 
&=\int_{0}^{(i-1)/n}\frac{n}{1+(1-\tfrac{1}{n})^{ny-i}}dy,
\end{align*}
which yields (\ref{Int1}). The proof of (\ref{Int2}) is similar, except that here we use the function   $f(2i-x)$, which is decreasing.

For (\ref{Int3}):
    \begin{align*}
        0\le \sum_{j=1}^{n}f(j)-\int_0^n f(x)dx
        &\le\sum_{J=1}^n ( 1\cdot (f(j)-f(j-1)))\\
        &\le f(n)\le 1.
    \end{align*}
    Now, subtracting $f(i)=1/2$ from all  sides to account for the missing $i$-th term in the first sum in (\ref{Int3}), we obtain the required estimate.
    
    \qed
\end{Proof}

\begin{lemma}
\label{prop:ord}
For $0 \leq i \leq n$:
\begin{description}
\item {1. }   $0\le\log\left(1+e^{i/n-1}\right)-\log\left(1+(1-\tfrac{1}{n})^{n-i}\right)\le\tfrac{1}{n}$
 \item {2. }  $-\tfrac{1}{n}\le\log\left(1+e^{i/n}\right)-\log\left(1+(1-\tfrac{1}{n})^{-i}\right)\le0$
 \item  {3. }  $-\tfrac{1}{n}\le\log\left(1+e^{1-i/n}\right)-\log\left(1+(1-\tfrac{1}{n})^{i-n}\right)\le0$.
\end{description}
\end{lemma}

\newpage

\begin{Proof}{} 
\begin{description}
\item{1. } A routine calculation yields:
\begin{align*}
    0 &\leq \log\left(1+e^{i/n-1}\right)-\log\left(1+(1-1/n)^{n-i}\right)
    =\log\left(\frac{1+e^{i/n-1}}{1+(1-1/n)^{n-i}}\right)\\
    &=\log\left(1+\frac{e^{i/n-1}-(1-1/n)^{n-i}}{1+(1-1/n)^{n-i}}\right)
    \leq \log\left(1+e^{i/n-1}-(1-1/n)^{n-i}\right)\\
    &\leq e^{i/n-1}-(1-1/n)^{n-i}
    \leq e^{i/n-1}-e^{\frac{i-n}{n-1}}
    \leq\left(i/n-1-\frac{i-n}{n-1}\right)e^{i/n-1}\\
    &\leq i/n-1-\frac{i-n}{n-1}=\frac{n-i}{n(n-1)}\leq \frac{1}{n}.
\end{align*}
\item{2. } Proved similarly to the preceding part.
\item{3. } Follows by replacing $i$ with $n-i$ in the previous part.
\end{description}

\qed
\end{Proof}

\begin{Proof}{ of Theorem \ref{th:location}}
\begin{description}
\item{1. }
The probability that edge $e_i$ arrives before edge $e_j$ is
\begin{align*}
\frac{p_i}{p_i+p_j},
\end{align*}
so that
\begin{align*}
E(L_n(i))&=n-\sum_{j=1, j\neq i}^n\frac{p_i}{p_i+p_j}
=n-\sum_{j \neq i}\frac{p(1-p)^{i-1}}{p(1-p)^{i-1}+p(1-p)^{j-1}}\\
&=n-\sum_{j \neq i}\frac{1}{1+(1-p)^{j-i}}.
\end{align*}
Also, for every $k$,
$$\frac1{1+(1-p)^k}+\frac1{1+(1-p)^{-k}}=1.
$$Therefore, using the change of variables $\ell=n-j+1$:
\begin{align*}
    &E(L_n(i))+E(L_n(n-i+1))\\
    &\hspace{10mm}=\left(n-\sum_{j \neq i}\frac{1}{1+(1-p)^{j-i}}\right)
    +   \left(n-\sum_{j \neq n-i+1}\frac{1}{1+(1-p)^{j-n+i-1}}\right)\\
    &\hspace{10mm}=2n+1-\sum_{j=1}^{n}\frac{1}{1+(1-p)^{j-i}}
    -\sum_{j = 1}^{n}\frac{1}{1+(1-p)^{j-n+i-1}}\\
    &\hspace{10mm}=2n+1-\sum_{j=1}^{n}\frac{1}{1+(1-p)^{j-i}}
    -\sum_{\ell = 1}^{n}\frac{1}{1+(1-p)^{i-\ell}}\\
    &\hspace{10mm}=2n+1-\sum_{j=1}^{n}1
    =n+1.
\end{align*}

\item{2. }
By (\ref{Int3})
\begin{align*}
E(L_n(i))=
n-\int_{0}^{1}\frac{n}{1+(1-1/n)^{ny-i}}dy+O(1).
\end{align*}
Employing the substitution $t=(1-1/n)^{ny-i}$, we obtain:
\begin{align*}
\int_{0}^{1}\frac{1}{1+(1-\tfrac{1}{n})^{ny-i}}dy
&=1-\frac{1}{\log \left((1-\tfrac{1}{n})^n\right)}\int_{(1-1/n)^{-i}}^{(1-1/n)^{n-i}}\frac{dt}{1+t}\\
&=1-\frac{\log\left(1+(1-\tfrac{1}{n})^{n-i}\right)-\log\left(1+(1-\tfrac{1}{n})^{-i}\right)}{\log \left((1-\tfrac{1}{n})^n\right)}.\\
\end{align*}
Noting that
\begin{equation}\label{eq6}
\frac{1}{\log\left((1-1/n)^n\right)}=-1+O(1/n),
\end{equation}
Lemma \ref{prop:ord}.1 and Lemma \ref{prop:ord}.2 give:
\begin{align*}
&\int_{0}^{1}\frac{1}{1+(1-1/n)^{ny-i}}dy\\
&\hspace{10mm}=1-\left(\log\left(1+e^{i/n-1}\right)-\log\left(1+e^{i/n}\right) +\,O({1}/{n})\right)\cdot\left(-1+O(1/n)\right)\\
&\hspace{10mm}=1+\log\left(1+e^{i/n-1}\right)-\log\left(1+e^{i/n}\right)+O(\tfrac{1}{n}).
\end{align*}
Hence:
\begin{align*}
   E(L_n(i))
   &=   n\left(
   1-\int_{0}^{1}\frac{1}{1+(1-1/n)^{ny-i}}dy
   \right)+O(1)\\
   &=n\left(\log\left(1+e^{i/n}\right)-\log\left(1+e^{i/n-1}\right)\right)+O(1).
\end{align*}
\end{description}
\qed
\end{Proof}

\begin{Proof}{ of Theorem~\ref{th:disr}}
\begin{description}
\item {1.}
The expected number of disruptions involving $i$ is given by:
\begin{align*}
E(D_n(i))
&=\sum_{j=1}^{i-1}
\frac{1}{1+(1-p)^{j-i}}
+
\sum_{j =i}^{n}
\frac{1}{1+(1-p)^{i-j}}-1/2.
\end{align*}
A routine change of variable yields:
\begin{align*}
E(D_n(n+1-i))
&=\sum_{j=1}^{n-i}
\frac{1}{1+(1-p)^{j-n-1+i}}
+\sum_{j =n-i+1}^{n}
\frac{1}{1+(1-p)^{n+1-i-j}}-1/2\\
&=E(D_n(i)).
\end{align*}
\item {2.} By (\ref{Int1}) and (\ref{Int2}):
\begin{align*}
\Bigg\vert E(D_n(i))&-
\int_{0}^{\frac{i-1}{n}}\frac{n}{1+(1-\frac{1}{n})^{ny-i}}dy
-
\int_{\frac{i}{n}}^{1}\frac{n}{1+(1-\frac{1}{n})^{i-ny}}dy\Bigg\vert
\leq 3/2.
\end{align*}
As in the proof of Theorem \ref{th:location}.2,  we employ the substitution $t=(1-1/n)^{ny-i}$ to obtain:
\begin{align*}
\int_{0}^{(i-1)/n}\frac{n}{1+(1-\frac1n)^{ny-i}}dy
&=i-1-\frac{n}{\log \left((1-\frac1n)^n\right)}
\int_{(1-1/n)^{-i}}^{(1-1/n)^{-1}}\frac{dt}{1+t}\\
&=i-1-\frac{n}{\log \left((1-\frac1n)^n\right)}
\cdot\left(\log\left(1+(1-\tfrac1n)^{-1}\right)\right.\\
&\hspace{3mm}\left.-
\log\left(1+(1-\tfrac1n)^{-i}\right)\right).
\end{align*}
Noting that
\begin{align*}
\log(2+1/(n-1))-\log2&=O(1/n),
\end{align*}
thus, by (\ref{eq6}) and Lemma \ref{prop:ord}.2
\begin{align*}
    &\int_{0}^{(i-1)/n}\frac{n}{1+(1-1/n)^{ny-i}}dy
    =    i-1-n\left(\log\left(1+e^{i/n}\right)-\log2\right)+O(1).
\end{align*}
Finally, employing the substitution $t=(1-1/n)^{i-ny}$, we obtain 
\begin{align*}
\int_{i/n}^{1}\frac{n}{1+(1-1/n)^{i-ny}}dy
&=n\int_{i/n}^{1}1dy
-
n\int_{i/n}^{1}\frac{(1-1/n)^{i-ny}}{1+(1-1/n)^{i-ny}}dy\\
&=n-i+\frac{n}{\log \left((1-1/n)^{n}\right)}
\int_{1}^{(1-1/n)^{i-n}}\frac{dt}{1+t}\\
&=n-i-\frac{n}{\log \left((1-1/n)^{-n}\right)}
\cdot \left.\log(1+t)\right|_{1}^{(1-1/n)^{i-n}}\\
&=n-i-\frac{n}{\log \left((1-1/n)^{-n}\right)}\\
&\hspace{16mm}\cdot
\left(\log\left(1+(1-1/n)^{i-n}\right)
-
\log 2\right).
\end{align*}

Thus, by Lemma \ref{prop:ord}.3
\begin{align*}
&\int_{i/n}^{1}
\frac{n}{1+(1-1/n)^{i-ny}}dy
=n-i-n\left(\log\left(1+e^{1-i/n}\right)-\log{2}\right)+O(1).
\end{align*}
The result now follows easily.
\end{description}

\qed
\end{Proof}

\begin{Proof}
{ of Theorem~\ref{maxB}}
\begin{description}
\item {1.}
The probability that $e_{i_1}$ will be the first edge to arrive  is $p_{i_1}/(1-q^n)$.
Suppose that $e_{i_1},e_{i_2},\ldots ,e_{i_{k-1}}$ have arrived. The probability that $e_{i_k}$ will be the next edge to arrive, is $p_{i_k}/(1-q^n-p_{i_1}-p_{i_2}-\cdots -p_{i_{k-1}})$. Thus 

\begin{equation}\label{311label0}
    \begin{aligned}
    P(B(i_1,\ldots ,i_n))
    &=\prod_{k=1}^n \frac{p_{i_k}}{1-q^n-p_{i_1}-\cdots -p_{i_{k-1}}}\\
    &=p^nq^{n(n-1)/2}
    \cdot\prod_{k=1}^n\frac1{1-q^n-p_{i_1}-\cdots -p_{i_{k-1}}}.
    \end{aligned}
    \end{equation}
In particular
\begin{equation}\label{thm310part1}
    \begin{aligned}
    P(\overline{B})&=p^n\prod_{k=0}^{n-1}\frac1{1-q^{n-k}},
    \qquad P(\underline{B})&=p^nq^{n(n-1)/2}\prod_{k=0}^{n-1}\frac1{1-q^{n-k}}. 
    \end{aligned}   
    \end{equation}
Consider the effect of transposing two consecutive edges $e_{i_k}$ and $e_{i_{k+1}}$ in the received order. By (\ref{311label0}) we easily obtain:
\begin{align*}
&\frac{P(B(e_{i_1},\ldots,e_{i_k},e_{i_{k+1}},\ldots,e_{i_n}))}{P(B(e_{i_1},\ldots,e_{i_{k+1}},e_{i_{k}},\ldots,e_{i_n}))}
=\frac{1-q^n-p_{i_1}-\cdots -p_{i_{k-1}}-p_{i_{k+1}}}{1-q^n-p_{i_1}-\cdots -p_{i_{k-1}}-p_{i_{k}}}.
\end{align*}
The right-hand side  is greater than $1$ if an only if $p_{i_{k+1}}<p_{i_{k}}$.

Take an arbitrary $B$. By a suitable sequence of swaps in  one direction  we obtain $\overline{B}$, and by a suitable sequence in the opposite direction we obtain $\underline{B}$. Thus,   
$$P(\overline{B})\geq P(B)\geq P(\underline{B}).$$
\item {2.} 
Replacing the logarithmic function by a MacLaurin series, we routinely find that:
\begin{align*}
&\sum_{i=0}^{n-1}-\log{\left(1-\left(1-\frac{1}{n}\right)^{n-i}\right)}
=\sum_{j=1}^{\infty}\frac{1-\left(1-\frac{1}{n}\right)^{nj}}{j}\cdot\frac{1}{\left(\frac{n}{n-1}\right)^{j}-1}.
\end{align*}
Note  that
\begin{align*}
\frac{1}{e^{j/(n-1)}-1} \leq \frac{1}{\left(\frac{n}{n-1}\right)^{j}-1} \leq \frac{1}{e^{j/n}-1} \leq \frac{n}{j}.
\end{align*}
Therefore:
\begin{align*}
\sum_{i=0}^{n-1}-\log{\left(1-\left(1-\frac{1}{n}\right)^{n-i}\right)}
&=
\sum_{j=1}^{\infty}\frac{1-\left(1-\frac{1}{n}\right)^{nj}}{j}\cdot\frac{1}{\left(\frac{n}{n-1}\right)^{j}-1}\\
&\leq 
\sum_{j=1}^{n-1}\frac{1-\left(1-\frac{1}{n}\right)^{nj}}{j}\cdot
\frac{1}{e^{j/n}-1}
+
O(1)\\
&=
n\sum_{j=1}^{n-1}\frac{1-\left(1-\frac{1}{n}\right)^{nj}}{j^2}\cdot
\frac{1}{\sum_{l=1}^{\infty} \frac{1}{l!}\left(\frac{j}{n}\right)^{l-1}}
+
O(1)\\
&=
n\sum_{j=1}^{n-1}\frac{1-\left(1-\frac{1}{n}\right)^{nj}}{j^2}
\cdot
\left(
\sum_{k=0}^{\infty}(-1)^k\left(\sum_{l=2}^{\infty} \frac{1}{l!}\left(\frac{j}{n}\right)^{l-1}\right)^k
\right)
+
O(1)\\
&\leq 
n\sum_{j=1}^{n-1}\frac{1-\left(1-\frac{1}{n}\right)^{nj}}{j^2}
\left(1-\frac{1}{2}\frac{j}{n}
\right)
+
O(1)\\
&=
n\sum_{j=1}^{n-1}\frac{1-\left(1-\frac{1}{n}\right)^{nj}}{j^2}
-
\frac{1}{2}\sum_{j=1}^{n-1}\frac{1-\left(1-\frac{1}{n}\right)^{nj}}{j}
+
O(1).
\end{align*}
From the other direction we similarly have:
\begin{align*}
\sum_{i=0}^{n-1}-\log{\left(1-\left(1-\frac{1}{n}\right)^{n-i}\right)}
&=
\sum_{j=1}^{\infty}\frac{1-\left(1-\frac{1}{n}\right)^{nj}}{j}\cdot\frac{1}{\left(\frac{n}{n-1}\right)^{j}-1}\\
&\geq 
\sum_{j=1}^{n-1}\frac{1-\left(1-\frac{1}{n}\right)^{nj}}{j}\cdot
\frac{1}{e^{j/(n-1)}-1}
+
O(1)\\
&=
\sum_{j=1}^{n-1}\frac{1-\left(1-\frac{1}{n}\right)^{nj}}{j}\cdot
\frac{1}{\sum_{l=1}^{\infty} \frac{1}{l!}\left(\frac{j}{n-1}\right)^l}
+
O(1)\\
&\geq 
(n-1)\sum_{j=1}^{n-1}\frac{1-\left(1-\frac{1}{n}\right)^{nj}}{j^2}
\left(1-\tfrac{1}{2}\tfrac{j}{n-1}
\right)
+
O(1)\\
&\geq 
n\sum_{j=1}^{n-1}\frac{1-\left(1-\frac{1}{n}\right)^{nj}}{j^2}
-
\frac{1}{2}\sum_{j=1}^{n-1}\frac{1-\left(1-\frac{1}{n}\right)^{nj}}{j}
+
O(1).
\end{align*}
Altogether:
\begin{align*}
\sum_{i=0}^{n-1}-\log{\left(1-\left(1-\frac{1}{n}\right)^{n-i}\right)}
&=
n\sum_{j=1}^{n-1}\frac{1-\left(1-\frac{1}{n}\right)^{nj}}{j^2}
-
\frac{1}{2}\sum_{j=1}^{n-1}\frac{1-\left(1-\frac{1}{n}\right)^{nj}}{j}
+
O(1)\\
&=
n\sum_{j=1}^{\infty}\frac{1-\left(1-\frac{1}{n}\right)^{nj}}{j^2}
-
\frac{1}{2}\sum_{j=1}^{n}\frac{1}{j}
+
O(1)\\
&=
n\left(\text{Li}_2(1)-\text{Li}_2\left(\left(1-\tfrac{1}{n}\right)^{n}\right)\right)-\tfrac{H_n}{2}+O(1).
\end{align*}

Thus:
\begin{align*}
    \log{\left(\prod_{i=0}^{n-1}(1-q^{n-i})\right)}
    &=n\text{Li}_2\left(e^{-1}\right)-
n\text{Li}_2\left(1\right)+\frac{1}{2}H_n+O(1)\\
    &=(\text{Li}_2\left(e^{-1}\right)-
\text{Li}_2\left(1\right))n+\log n/2+O(1)\\
    &\simeq -1.236n+\log n/2+O(1).
\end{align*}
By (\ref{thm310part1}):
\begin{align*}
P(\overline{B})&=
\frac{p^n}{\prod_{i=0}^{n-1}(1-q^{n-i})}
=\Theta\left({e^{\left({\pi^2}/{6}-\text{Li}_2\left(e^{-1}\right)\right)n}}/{n^{n-1/2}}\right)
\simeq \Theta\left({e^{1.236n}}/{n^n}\right).
\end{align*}
\item {3.} Follows from  parts 2 and 4. 
\item {4.}
By (\ref{thm310part1}):
\begin{align*}
P(\underline{B})/P(\overline{B})&=q^{n(n-1)/2}.
\end{align*}
Now
\begin{align*}
\log q^{n(n-1)/2}&=\frac{n(n-1)}{2}\log\left(1-\frac{1}{n}\right)
=\frac{-n(n-1)}{2}\left(\frac{1}{n}+\frac{1}{2n^2}+O\left(\frac{1}{n^3}\right)\right)\\
&=\frac{-n}{2}+\frac{1}{4}+O\left(\frac{1}{n}\right),
\end{align*}
so that
\begin{align*}
    P(\underline{B})/P(\overline{B})
    =e^{{-n}/{2}+{1}/{4}}\left(1+o(1)\right).
\end{align*}
\end{description}

\qed
\end{Proof}

\section{Technical Issues and Conclusion}\label{technicalities}
In this section we will delve deeper into some of the technical aspects of our solution.
First, we discuss the PPM method. Second, we discuss the assumptions made in Algorithms~\ref{Algorithm1} and \ref{Algorithm2}, presented in Section \ref{secAlgorithm}, and explain why they are reasonable. 

\subsection{The PPM Method}
The main idea of PPM was explained in Section \ref{subsection:PPM_intro}. The main advantage of this method for performing an IP traceback is its efficacy, which is expressed by two main features:  
\begin{enumerate}
    \item Low additional overhead on the packets' size (i.e., on the communication resources).
   The additional overhead is necessary for the packet markings. There are several approaches as to how to encode the edges efficiently in the header of the packet (using the Chinese Remainder Theorem \cite{CRT} and other methods \cite{AKSW,ENC_EX2}). 
    \item Low additional processing time for the victim and the routers.
\end{enumerate}

The main difference between the various PPM algorithms is in the reconstruction algorithm executed by the victim. The routers, on the other hand, preform a marking algorithm, which is mostly the same.

\subsection{The Assumptions in Our Algorithms}
Our algorithms are simple and can be adapted to most variants for our problem. They rely on the following assumptions (which are common to PPM algorithms for IP trackback):
\begin{enumerate}
    \item The path from the attacker to the victim is fixed, and is not under the attacker's control. This assumption about the internet network dynamics is very weak. In fact, forwarding tables do not update too frequently, so the paths between computers do not change frequently.
    \item The victim does not need to know the network structure or the routers' marking algorithm. On the other hand:
    
    \begin{enumerate}
        \item He knows the marking probability $p$, required for the decision when to stop collecting edges in Algorithm~\ref{Algorithm2}. (Note that Algorithm \ref{Algorithm1} does not require the value of $p$.)
        \item He is able to verify that a subpath is a full subpath. This ability can be implemented easily, for example by commands such as Linux traceroute \cite{traceroute}. Obviously, there are other methods that are much more cost-effective. 
    \end{enumerate}
\end{enumerate}

\section{Conclusion and Future Work}\label{secConclusion}

\subsection{The Advantages of Our Algorithms}
Our algorithms have several advantages: 
    \begin{enumerate}
        \item The speed of reconstruction of the attack path: Other algorithms,  such as SWKA \cite{AKSW} and S\&S \cite{SS2},  use a fixed parameter for the termination time of the algorithm (a predefined number of packets that should be received by the victim) in order to guarantee the correctness of the reconstructed graph with only small error probability. 
        Note that, as the length of the path increases, traditionally this termination time grows too, so for achieving reasonable accuracy, this parameter needs to depend on the maximum possible length of the attack path (the worst case). 
        In contrast, our algorithms use a dynamic termination time. The stopping time is determined by the actual length of the subpath the victim has. This dynamic time allows in some cases the graph to be reconstructed much faster, which results in improved victim's survivability.
    \item Both our algorithms need very little information as input. In fact, both do not need $n$, the distance of the attacker from the victim, as input. Moreover, Algorithm~\ref{Algorithm1} does not even need the marking probability $p$.
        \item Algorithm~\ref{Algorithm2} may be adapted by changing the parameter $\eps$.
        This enhancement can be exploited in a proper balance between the defense needs (whose importance changes during the attack) and the desire to allow legitimate users transmit packets without disruptions.
    \end{enumerate}

\subsection{Adaptation to Multiple Attackers with Shared Edges}
In \cite{AKSW} it is explained that edge sampling algorithms can efficiently discern multiple attackers ``because attackers from different sources produce disjoint edges in the tree structure used during reconstruction". This assumption is not completely accurate, due to the case of shared edges between paths of different attackers, as shown in \cite{LWW,Kiremire2012}. Yet, this approach approximates the attack graph. This approximation approach can easily integrate with our algorithms. 
Later, in \cite{SS2}, it has been suggested how one can generalize PPM algorithms to the case of multiple attackers when there are shared edges in distinct attack paths. The method is based on counting the marks generated by each attack path separately. This technique is less effective but more accurate. It will be interesting to compare our approach in the case of shared edges to this approach.

\appendices

\newpage

\section{Other Simulation Results}\label{OtherSimResults}

In the simulation described in Section \ref{simulations}, we have gathered additional data regarding the algorithms and the  paths   encountered during the run.

 When either one of the algorithms SWKA and S\&S comes to an end, we get one of the following: 
 
 (i) The full attack path. 
 
 (ii) A full subpath, but not the full attack path -- one or more of the last edges are missing. 
 
 (iii) A collection of edges, not constituting a full subpath. Namely, we have a ``hole" in the path. 

In Table~\ref{table3}, we present data regarding the frequency of each outcome for both algorithms. The data for SWKA appears in the second column, and for S\&S -- in the third. In the second line of the table, we present the percent of times we got option (i), in the third -- option (ii), and in the fourth -- option (iii). (Of course, the second line gives also the success rate of the algorithms.) 

Note that most failures of the two algorithms are not due to having holes. Thus, usually, when they fail, it is impossible to know that a false path was obtained.

 \begin{table}[ht]
\renewcommand{\arraystretch}{1.3}
\label{table3}
\centering
\begin{tabular}{|c||c|c|}
\hline
 & SWKA & S\&S \\
\hline
\makecell{(i) Full\\Attack Path}
& $0.77$ & $0.86$ \\ \hline
\makecell{(ii) Full Subpath, but\\Not a Full Attack Path}
& $0.20$ & $0.12$\\ \hline
\makecell{(iii) Not a Full\\Subpath}
& $0.03$ &$0.02$\\ \hline
\end{tabular}
\caption{Additional statistics for SWKA and S\&S.}
\end{table}

In Table~\ref{table4}, we present statistics on the full subpaths obtained during the run, when we continue until all edges have been collected.
The first column of the table lists the full subpath lengths one may encounter in the process. 
For each such length, the second column provides the probability of Algorithm~\ref{Algorithm1} to return a full subpath of that length, and the third column -- the probability of encountering such a full subpath during the reconstruction of $P$.

 \begin{table}[ht]
\renewcommand{\arraystretch}{1.3}
\label{table4}
\centering
\begin{tabular}{|c||c|c|}
\hline
\makecell{Full Subpath\\ Length} &\makecell{Probability of\\Algorithm~\ref{Algorithm1} to\\  Return Such Length}& \makecell{Probability to\\Get Such a Length\\ in an Iteration}\\
\hline
$2$ & $8.0\cdot 10^{-3}$ & $8.0\cdot 10^{-3}$ \\ \hline
$3$ & $1.0\cdot 10^{-3}$ & $1.6\cdot 10^{-3}$ \\ \hline
$4$ & $8.0\cdot 10^{-3}$ & $4.1\cdot 10^{-4}$ \\ \hline
$5$ & $1.0\cdot 10^{-4}$ &  $1.5\cdot 10^{-4}$ \\ \hline
$6$ & $4.7\cdot 10^{-5}$ & $6.5\cdot 10^{-5}$ \\ \hline
$7$ & $2.4\cdot 10^{-5}$ & $3.0\cdot 10^{-5}$ \\ \hline
$8$ & $1.4\cdot 10^{-5}$ & $1.7\cdot 10^{-5}$ \\ \hline
$9$ & $1.1\cdot 10^{-5}$ & $1.3\cdot 10^{-5}$ \\ \hline
$10$ & $6.8\cdot 10^{-6}$ & $8.4\cdot 10^{-6}$ \\ \hline
$11$ & $6.1\cdot 10^{-6}$ & $6.6\cdot 10^{-6}$ \\ \hline
$12$ & $8.4\cdot 10^{-6}$ & $9.1\cdot 10^{-6}$ \\ \hline
$13$ & $8.9\cdot 10^{-6}$ & $1.0\cdot 10^{-5}$ \\ \hline
$14$ & $1.1\cdot 10^{-5}$ & $1.2\cdot 10^{-5}$ \\ \hline
$15$ & $1.5\cdot 10^{-5}$ & $1.7\cdot 10^{-5}$ \\ \hline
$16$ & $2.6\cdot 10^{-5}$ & $2.8\cdot 10^{-5}$ \\ \hline
$17$ & $3.9\cdot 10^{-5}$ & $4.4\cdot 10^{-5}$ \\ \hline
$18$ & $7.4\cdot 10^{-5}$ & $8.1\cdot 10^{-5}$ \\ \hline
$19$ & $1.6\cdot 10^{-4}$ & $1.7\cdot 10^{-4}$ \\ \hline
$20$ & $4.1\cdot 10^{-4}$ & $4.5\cdot 10^{-4}$ \\ \hline
$21$ & $1.2\cdot 10^{-3}$ & $1.3\cdot 10^{-3}$ \\ \hline
$22$ & $4.1\cdot 10^{-3}$ & $4.5\cdot 10^{-3}$ \\ \hline
$23$ & $1.7\cdot 10^{-2}$ & $1.9\cdot 10^{-2}$ \\ \hline
$24$ & $9.7\cdot 10^{-2}$ & $1.1\cdot 10^{-1}$ \\ \hline
$25$ & $0.87$ & $1.00$ \\ \hline
\end{tabular}
\caption{More statistics on the full subpaths encountered in the experiment.}
\end{table}

Note that the probabilities in the second column are of pairwise disjoint events (and so sum up to~1), while those in the third are not.

\end{document}